\documentclass[11pt]{article}
\newlength{\vshift}
\newlength{\hshift}
\usepackage{amssymb}
\usepackage{amsmath}
\usepackage{amsthm}
\usepackage{amsopn}
\def\diff{\textrm{d} }
\def\nn{\nonumber }
\def\la{\lambda}

\def\ka{\kappa}
\def\de{\delta}
\def\be{\beta}

\def\al{\alpha}
\def\si{\sigma}

\def\shouldid{\stackrel{!}{=}}

\def\hxi{\hat \xi}
\def\hdi{\hat D}
\def\hom{\hat \omega}

\def\xx{{\hat x}}

\def\hp{\hat{\partial}}
\def\p{\partial}

\def\hp{\hat{\partial}}

\def\lb{\lbrack}
\def\rb{\rbrack}
\def\pat{\partial}
\def\dw{\stackrel{\star}{\wedge}}

\providecommand{\href}[2]{#2}

\begin{document}

\begin{titlepage}
\rightline{LMU-TPW 2004-08}
\rightline{MPP-2004-109}

\vspace{4em}
\begin{center}

{\Large{\bf A symmetry invariant integral on $\kappa$-deformed
    spacetime}}

\vskip 3em

{{\bf  Lutz M\"oller}}

\vskip 1em

Universit\"at M\"unchen, Fakult\"at f\"ur Physik\\
        Theresienstr.\ 37, D-80333 M\"unchen\\[1em]

Max-Planck-Institut f\"ur Physik\\
        F\"ohringer Ring 6, D-80805 M\"unchen\\[1em]
 \end{center}

\vspace{2em}

\begin{abstract}
In this note we present an approach using both constructive 
and Hopf algebraic methods
to contribute to the not yet fully satisfactory definition of an
integral on $\ka$-deformed spacetime. The integral presented here is based on the inner product
of differential forms and it is shown that this integral is explicitly
invariant under the deformed symmetry structure. 

\end{abstract}
\vskip 0.5cm
\qquad\hspace{2mm}\scriptsize{eMail: lutz@lutzmoeller.de}
\vfill

\end{titlepage}\vskip.2cm

\newpage
\setcounter{page}{1}

\section{Introduction}
\label{A}
The $\ka$-deformed spacetime is frequently discussed as one of the
most attractive models for a noncommutative (NC) spacetime, since it can be
endowed with a
deformed symmetry structure \cite{lukrue1}, \cite{lukrue2},
\cite{majrue}, \cite{local}, \cite{gac1}. Recently, there has been substantial progress in
reinterpreting traditional algebraic-geometric concepts in the
NC setting \cite{f1},  \cite{f2}, \cite{f3}. However, many problems persist.
In this note we present constructive methods
combined with some Hopf algebraic concepts
to contribute to the not yet fully satisfactory definition of an
integral \cite{f1}, \cite{gac2} for the $\ka$-deformed spacetime. The integral presented here is based on the inner product
of differential forms. Thereby this note continues the recent
definition of an $n$-dimensional calculus of one-forms for an
$n$-dimensional $\ka$-deformed space \cite{f3}. It is shown that this integral is explicitly
invariant under the deformed symmetry structure. The drawback
is that this integral is not cyclic (at least not at first sight),
therefore it seems not to be
useful for gauge theory. 

The work is a consequently new approach in a series of previous attempts of other
authors to construct an integral invariant under ·-deformed symmetry transformations.
Previous works have focused for example \cite{9} on the deformed Fourier theory, to build a
wave-packet using an integration invariant under the deformed action of Lorentz transformations.
The same paper proposes also a left invariant and a right invariant integration
measure; a similar construction is derived in  \cite{10}, analysing different ordering procedures
and the action of translation generators. Again, the construction in  \cite{11} is based on a
general analysis of an equation of motion with an infinite number of derivatives. The four
approaches of those three papers just mentioned differ fundamentally in derivation and
result with the results of this paper at hand. It is of course not claimed here that the best
or even most useful definition of an invariant (though not cyclic) integral on $\kappa$-deformed
space has been constructed. The integral of this paper is actually an inner product; this allows
a very rigid Hopf-algebraic treatment, but the conceptual limitations of this approach
for a general notion of an integral are obvious. Still, the results of this paper are rather
unambiguous (except up to ordering procedures) and to all orders.

This note is organised as follows: in section \ref{abs} we recall the
most important definitions of \cite{f1} and \cite{f3}. In section
\ref{formint} we define the integral as an inner product of
differential forms. In section \ref{intinv} we show that this integral is
invariant under the deformed symmetry structure. In appendix
\ref{cyclic} we recall the definition of a cyclic integral from
\cite{f1} and in \ref{hermitian} we recall the definition of hermitian
symmetry generators discussed as well in \cite{f1}. This note should
be read as a juxtaposition of two different approaches, the one in the
main part of this note and the other in the appendix. We conclude that
at present neither of the two approaches is without fail,
they will have to be combined in the future in an ingenious way. 

%
%
%
%

\section{Abstract $\ka$-Euclidean algebra}
\label{abs}
In this note we use the conventions of \cite{f1} and
\cite{f3}. The $\ka$-deformed spacetime (a Lie algebra space with
structure constants
$C^{\mu\nu}_\la=(a^\mu\de^\nu_\la-a^\nu\de^\mu_\la)$) is discussed in its Euclidean
version for simplicity, the characteristic vector $a^\mu$ is aligned
with the $n$th direction, therefore 
\begin{equation}
\label{nc14}
\lb \xx^n, \xx^j\rb =ia\xx^j, \qquad \lb \xx^i,\xx^j\rb=0, \quad
\forall i,j \in \{1, \dots , n-1\}.
\end{equation}
There is a deformed Poincar\'e symmetry structure on this spacetime:
\begin{eqnarray}
\label{k7}
\lb \hat{M}^{rs}, \hat{x}^n\rb &=& 0,\qquad\qquad\quad\quad\hspace{2mm}
\lb \hat{M}^{rs}, \hat{x}^j\rb = \de^{rj}\hat{x}^s -\de^{sj}\hat{x}^r,\nonumber \\
\lb \hat{N}^{l}, \hat{x}^n\rb &=& \hat{x}^l+ia \hat{N}^{l} , \quad\qquad\hspace{2mm}
\lb \hat{N}^{l}, \hat{x}^j\rb =  - \de^{lj}\hat{x}^n -ia \hat{M}^{lj},\\
\lb\hat{D}_n, \hat{x}^n \rb&=&
\sqrt{1-a^2\hat{D}_\mu\hat{D}_\mu}, \quad \hspace{1mm}\lb\hat{D}_n, \hat{x}^j \rb= ia\hat{D}_j,\nonumber \\
\lb\hat{D}_i, \hat{x}^n \rb&=& 0, \qquad \qquad \qquad\quad \hspace{1mm}\lb\hat{D}_i, \hat{x}^j \rb= \de^j_i\Big(-ia\hat{D}_n+
\sqrt{1-a^2\hat{D}_\mu\hat{D}_\mu}\Big),\nn \end{eqnarray}
however, this deformed symmetry is undeformed in the algebraic sector.
There are two $n$-dimensional bases of differential one-forms. One of them  transforms
vector-like under rotations, but has difficult commutation relations
with coordinates ($\mu,\nu=1,\dots,n$)
\begin{equation}
\label{d4}
 \lb \hxi^\mu, \xx^\nu\rb= ia (\de^{\mu n}\hxi^\nu
 -\de^{\mu\nu}\hxi^n)+(\hxi^\mu\hdi_\nu+\hxi^\nu\hdi_\mu
 -\de^{\mu\nu}\hxi^\rho\hdi_\rho)\frac{1-\sqrt{1-a^2\hdi_\si
    \hdi_\si}}{\hdi_\la \hdi_\la},
\end{equation} 
and one which commutes with coordinates, 
but transforms under rotations as\footnote{The derivatives $\widetilde{\p}_j$ used in (\ref{d67}) are defined as 
\begin{eqnarray}
\label{d68}
\lb \widetilde{\p}_i, \xx^n \rb &=&ia\widetilde{\p}_i,\qquad\qquad\lb \widetilde{\p}_n, \xx^n \rb =\frac{-ia^3\widetilde{\p}_s\widetilde{\p}_s\widetilde{\p}_n+\sqrt{1-a^2\widetilde{\p}_\mu\widetilde{\p}_\mu}}{1-a^2\widetilde{\p}_k\widetilde{\p}_k} ,\nn\\
\lb \widetilde{\p}_i, \xx^j \rb &=&\de^j_i, \qquad\qquad\quad
\lb \widetilde{\p}_n, \xx^j \rb =-ia\widetilde{\p}_j\frac{-ia\widetilde{\p}_n+\sqrt{1-a^2\widetilde{\p}_\mu\widetilde{\p}_\mu}}{1-a^2\widetilde{\p}_k\widetilde{\p}_k} .
\end{eqnarray}}
\begin{eqnarray}
\label{d67}
\lb \hat{M}^{rs}, \hom^n\rb&=&0,\qquad\quad 
\lb \hat{M}^{rs}, \hom^j\rb= \de^{rj}\hom^s-\de^{sj}\hom^{r},\nonumber\\
\lb \hat{N}^{l}, \hom^n\rb&=&
\hom^l\sqrt{1-a^2\widetilde{\p}_\mu\widetilde{\p}_\mu} +ia
(\hom^l\widetilde{\p}_n-\hom^n\widetilde{\p}_l),\\
\lb \hat{N}^{l}, \hom^j\rb&=&-\de^{lj}
\hom^n\sqrt{1-a^2\widetilde{\p}_\mu\widetilde{\p}_\mu} +ia
(\hom^l\widetilde{\p}_j-\hom^j\widetilde{\p}_l).\nonumber
\end{eqnarray}
Since the calculus of one-forms is $n$-dimensional, we have the
ordinary deRham differential calculus at our disposition. Especially,
there is one $n$-form,  the volume form
\begin{equation}
\label{d40a}
\lb \hat{M}^{rs},\hom^{1} \dots
 \hom^{n} \rb=0,\qquad\qquad\lb \hat{N}^{l},\hom^{1} \dots
 \hom^{n} \rb=-ia(n-1)\hom^{1} \dots
 \hom^{n}\hp_l.
\end{equation}
There are other 
derivatives with useful properties. The simplest derivatives are
 \begin{eqnarray}
\label{1.8}
\lb\hp _n, \hat x^j\rb &=& 0,\qquad
\lb\hp _n, \hat x^n\rb = 1,\nonumber\\
\lb\hp _i, \hat x^j\rb &=& \delta^j_i,\qquad
\lb\hp _i, \hat x^n\rb = ia \hp _i.
\end{eqnarray}
All these symmetry generators and forms etc. cannot only be defined in
the abstract algebra, but can be realised on ordinary functions,
replacing ordinary pointwise multiplication with the
$\star$-product. In the symmetric ordering, the summed-up $\star$-product for
$\ka$-deformed space reads
\begin{eqnarray}
\label{nc39}
 f(x) \star  g(x) &=& \lim_{\substack{y\to x \\ z \to x}}\exp \Bigg( x^j \pat_{y^j}
   \left( e^{-i\hbar a\pat_{z^n}} \frac{-i\hbar a\pat_n}{e^{-i\hbar a\pat_n}-1}\frac{e^{-i\hbar a\pat_{y^n}}-1}{-i\hbar a\pat_{y^n}} - 1 \right)\nonumber\\
  & & \qquad\qquad + x^j \pat_{z^j}
   \left( \frac{-i\hbar a\pat_n}{e^{-i\hbar a\pat_n}-1}\frac{e^{-i\hbar a\pat_{z^n}}-1}{-i\hbar a\pat_{z^n}} - 1 \right)
   \Bigg) f(y)g(z).
\end{eqnarray}
All symmetry generators can be represented (on ordinary
functions with this $\star$-product), e.g.
\begin{eqnarray}
\label{re2d}
 N^{\star l} f(x) & =& \Big( x^l\pat_n - x^n\pat_l
+ x^l\pat_\mu\pat_\mu \frac{e^{ia \pat_n}-1}{2\pat_n} - x^\mu\pat_\mu\pat_l
\frac{e^{ia\pat_n} -1- ia\pat_n}{ia\pat_n^2} \Big) f(x),\nonumber\\
M^{\star rs} f(x)& =& (x^s\pat_r-x^r\pat_s)f(x),\nn\\
D^\star _n
  f(x)&=&\Big(\frac{1}{a}\sin(a\partial_n)-\frac{1}{ia\partial_n\partial_n}
  \partial_k\partial_k(\cos(a\partial_n)-1)\Big)f(x),\\
D^\star _j
  f(x)&=& \partial_j \Big(\frac{e^{-ia\partial_n}-1}{-ia\partial_n}\Big) f(x).\nonumber
\end{eqnarray}
Also the forms can be represented, e.g. $\hom^\mu\rightarrow
\mathrm{d}x^\mu$. For further details  we
refer to \cite{f1} and
\cite{f3}. 

\section{Integration of forms}\label{formint}
Integrals for physical actions may be formulated as inner
products of forms. In commutative physics, actions are often written in
terms of the inner product of two differential $r$-forms $\psi$ and $\phi$, using the Hodge-$*$
operator\footnote{Note the different symbols for the $\star$-product and the
Hodge-$*$.}. In the case of an $n$-dimensional commutative manifold 
the Hodge-$*$ is  defined on an $r$-form\footnote{Conventions are
  according to \cite{nakahara}.}
\begin{equation}
\label{in69}
\phi = \frac{1}{r!}
\phi_{\mu_1\dots\mu_r}\textrm{d}x^{\mu_1}\wedge \dots\wedge \textrm{d}x^{\mu_r},
\end{equation}
as 
\begin{equation}
\label{in70}
*\phi = \frac{\sqrt{\det g}}{r!(n-r)!}
\phi_{\mu_1\dots\mu_r}\epsilon^{\mu_1\dots\mu_r}_{\ \ \
  \nu_{r+1}\dots \nu_{n}}\textrm{d}x^{\nu_{r+1}}\wedge \dots\wedge \textrm{d}x^{\nu_n}.
\end{equation}
Here $g$ is the metric defined on the commutative manifold. Recall the
identities $*1=\sqrt{\det g}\hspace{2mm}\textrm{d}^nx$ and $*^2\phi
= (-1)^{r(n-r)}\phi$.
The inner product of two $r$-forms is then the integral over the full spacetime times a measure:
\begin{equation}
\label{in71}
(\psi, \phi)= \int \psi \wedge *\phi=\frac{1}{r!}\int
\textrm{d}^n x \sqrt{\det g}\hspace{2mm}\psi_{\mu_1\dots\mu_r}\phi^{\mu_1\dots\mu_r}.
\end{equation}
Most physically relevant actions such as the Yang-Mills action and the minimally
gauge-coupled fermionic action  can be formulated in such a language. Locally, gauge potentials are Lie algebra-valued one-forms
$A^0=iA^0_\mu \textrm{d}x^\mu$. The field strength
$F^0_{\mu\nu}$ are components of a Lie algebra-valued two-form,
$F^0=\textrm{d}A^0+A^0\wedge A^0=iF^0_{\mu\nu}\textrm{d}x^\mu\wedge\textrm{d}x^\nu$,
fulfilling the Bianchi identity d$F^0 +
F^0\wedge A^0+A^0\wedge F^0=0$. 

To be more specific, the
Yang-Mills action is of the form:
\begin{eqnarray}
\label{in72}
( F^0, F^0)&=& \textrm{Tr}\int
(iF^0_{\mu\nu}\textrm{d}x^\mu\wedge \textrm{d}x^\nu) \wedge
*(i F^0_{\rho\si}\textrm{d}x^\rho\wedge \textrm{d}x^\si)\nn\\
&=&-\frac{1}{2}\textrm{Tr}\int
\textrm{d}^n x \sqrt{\det g}\hspace{2mm}F^0_{\mu\nu}F^{0\mu\nu}.
\end{eqnarray}
Since the
$\ka$-deformed space in our ansatz is considered to be flat, it is sufficient  to
treat spinor fields as fields of form degree zero. We do not need to
consider the Dirac derivative as the sum of two Dirac operators acting on
the two spin bundles which make up the exterior bundle. Therefore
spinorial actions fit into the framework.

In analogy to this phrasing of commutative actions, we now want to formulate NC field theories in the 
language of forms. Of course, we need to replace all point-wise products with
$\star$-products. From equation (\ref{in72}) we see that we also need a suitable set of differential forms which can be combined into a
volume form.
For example in the Yang-Mills action, one of the two two-forms has to be
commuted through the field-components $F^0_{\mu\nu}$ in order to be
combined into a volume form. Therefore the 
frame one-forms  $\hat{\omega}^\mu$ which have been defined such  that they commute with
functions (aand in the $\star$-product setting can be
identified with ordinary one-forms
 $\hat{\omega}^\mu\rightarrow =\textrm{d}x^\mu$) do the job. 

This  means that the NC Yang-Mills action can be written in
the following way, commuting frame one-forms to the left
($\dw$ is simultaneously a wedge and a $\star$-product):
\begin{eqnarray}
\label{in75}
(F,F)&=& \textrm{Tr}\int
(iF_{\mu\nu}\omega^\mu \omega^\nu) \dw
*(i F_{\rho\si}\omega^\rho \omega^\si)\nn\\
&=&-\frac{1}{2}\textrm{Tr}\int
\omega^{\mu_1}\dots\omega^{\mu_n} \hspace{2mm}F_{\mu\nu}\star \big(\sqrt{\det g}F^{\mu\nu}\big).
\end{eqnarray}

The Hodge-$*$ applied to the field strength tensor on the right (or in
general to the second differential form) is proportional to 
$\sqrt{\det g}$. The authors of \cite{syko1} have found that
$\sqrt{\det g}$ can be identified with  the  measure $\mu$ (introduced
in \cite{f1}, discussed  once more in appendix \ref{cyclic}). The measure $\mu$ is
the Pfaffian of the NC structure, given by ($C^{\mu\nu}_\la$ are the
Lie algebra
structure constants of the NC space)
\begin{equation}
\label{in73}
\mu=\det\hspace{0.1mm}^{-\frac{1}{2}}(x^\la
C^{\mu\nu}_\la)=\frac{1}{n!2^n}\epsilon_{\mu_1 \mu_2\dots \mu_{2n}} (x^\la
C^{\mu_1\mu_2}_\la)\dots  (x^\la
C^{\mu_{2n-1}\mu_{2n}}_\la).
\end{equation}
Since $x^\la
C^{\mu\nu}_\la$ is zero at the origin and not invertible there, the origin has to be
excluded for defining $\mu$.
Defining \cite{syko2} in the abstract algebra a radius $\hat{r}$ in the $(n-1)$-dimensional
subspace as $\hat{r}=\sqrt{\sum_{i=1}^{n-1}\xx^i\xx^i}$,  the
derivations $\hat{r}^j\hp_j$ and $\hp_n$ have ordinary Leibniz rules
(since $\xx^j \hat{f}(\xx) = (e^{-ia\hp_n}\hat{f}(\xx))\xx^j$). These derivations are identical to the commutative $r^j
\pat_j$ and $\pat_n$. These commutative derivations
can be used to construct a commutative metric 
\begin{equation}
\label{in74}
g=r^{-2}\sum_{i=1}^{n-1}(\textrm{d}x^i)^2+
(\textrm{d}x^n)^2=(\textrm{d}\ln r)^2+\textrm{d}\Omega^2_{n-2}+
(\textrm{d}x^n)^2,
\end{equation}
with $\textrm{d}\Omega^2_{n-2}$ the $(n-2)$-dimensional spherical volume element.
Therefore $\sqrt{\det g}= r^{-(n-1)}=\mu$ (more exactly, $\sqrt{\det
  g}=\mu_2$, cp. appendix  \ref{cyclic}). 

The measure $\sqrt{\det g}=\mu$ 
appears as part of the action of the Hodge-$*$ ($\epsilon$-tensor is
as usual fully antisymmetric)
\begin{equation}
\label{in76}
*(i F_{\rho\si}\omega^\rho \omega^\si)= \frac{\mu}{2!(n-2)!}
F_{\al\be}\epsilon^{\al\be}_{\ \ \
  \nu_{3}\dots \nu_{n}}\omega^{\nu_{3}} \dots\omega^{\nu_n}.
\end{equation}
If $\mu$ should play the role of a measure as in the appendix \ref{cyclic}, it should multiply the
volume element. It can be extracted from the second
$\star$-multiplicant because of its properties, $x^j\p_j\mu =-(n-1)\mu$, $\p_n\mu=0$. 
This leaves additional derivatives $\pat_n$ acting on the two
differential forms. We expand up to second order (for two arbitrary $r$-forms $\psi$ and $\phi$):
\begin{eqnarray}
\label{in77}
 \psi\star(\mu\phi)&=&\mu \psi\phi +\frac{ia}{2}\mu\big(\pat_n \psi x^j \pat_j\phi-x^j
\pat_j \psi \pat_n \phi\big)-\frac{ia}{2}(n-1)\mu\pat_n \psi \phi\nn\\
&&-\frac{a^2}{8}\mu \big(\pat^2_n\psi x^j x^k\pat_j\pat_k\phi -
x^j\pat_j\pat_n \psi x^k\pat_k \pat_n \phi +  x^j x^k\pat_j\pat_k \psi \pat^2_n
\phi\big)\nn\\
&&+\frac{a^2}{4}(n-1)\mu\big(\pat_n^2 \psi x^j\pat_j \phi-x^j\pat_j\pat_n\psi
\pat_n \phi\big)\\
&&-\frac{a^2}{8}(n-1)n\mu\pat_n^2 \psi \phi+\frac{a^2}{12}(n-1)\mu\pat_n^2 \psi \phi-\frac{a^2}{12}(n-1)\mu\pat_n \psi\pat_n \phi+\dots\hspace{2mm}.\nn
\end{eqnarray}
Under an integral allowing partial integration (Stokes' law), 
the derivatives $\pat_n$ can be combined into one derivative operator
($\pat_n$ commutes with the 
$\star$-product and $\mu$), which we call $K$:
\begin{equation}
\label{in78}
\int \textrm{d}^nx \hspace{2mm}\psi\star(\mu \phi) =\int \textrm{d}^nx \mu
\hspace{2mm}\psi \star (K \phi).
\end{equation}
Up to second order we find:
\begin{eqnarray}
\label{in79}
K&=&
1+\frac{ia}{2}(n-1)\pat_n-\frac{a^2(n-1)(n-2)}{8}\pat_n^2-\frac{a^2}{12}(n-1)\pat^2_n
+\dots\nn\\ &=&\big(1+\frac{ia}{2}\pat_n-\frac{a^2}{12}\pat^2_n-\dots\big)^{n-1}=\big(\frac{-ia\pat_n}{e^{-ia\pat_n}-1}\big)^{n-1}.
\end{eqnarray}
The reason for having identified an expansion up to second
order with an all orders expression will become clear in the next section. Continuing
the formulation of the action in terms of forms we will rediscover the
derivative operator $K$ from an entirely
different argument.  

Thus, we have constructed a version of the integral, in which the measure
function appears naturally outside of the $\star$-product (using $\omega^1\dots \omega^n=\diff^n x$):
\begin{eqnarray}
\label{in80}
(F, F)&=& \textrm{Tr}\int
(iF_{\mu\nu}\omega^\mu\omega^\nu) \star
*(i F_{\rho\si}\omega^\rho\omega^\si)\nn\\
&=&\textrm{Tr}\int (iF_{\mu\nu}\omega^{\mu}\omega^{\nu})\star\big(\frac{\mu}{2!(n-2)!}
F_{\al\be}\epsilon^{\al\be}_{\ \ \
  \nu_{3}\dots \nu_{n}}\omega^{\nu_{3}} \dots\omega^{\nu_n}\big)\\
&=&-\frac{1}{2}\textrm{Tr}\int
\omega^{1}\dots\omega^{n}\mu\hspace{2mm}F_{\mu\nu}\star\big(K
F^{\mu\nu}\big)=-\frac{1}{2}\textrm{Tr}\int
\diff^n x\mu\hspace{2mm}F_{\mu\nu}\big(K F^{\mu\nu}\big),\nn
\end{eqnarray}
since $\mu$ allows to eliminate one $\star$-product (cp. appendix
\ref{cyclic}). Still, we have to understand better the role of the
operator $K$. This is the content of the next section.

%
%
%
%
%
%

\section{Invariance of the integral}
\label{intinv}
The definition of the integral in \cite{f1} has been
motivated to achieve the trace property,  invariance
under $SO_a(n)$ rotations has not been a guiding principle in the
construction. 
In contrast we will now investigate the integral of inner product of
forms, and will find that this is
$SO_a(n)$-invariant by definition. 
 Since $SO_a(n)$ is a Hopf
algebra, we have to change the usual notion of invariance used in the
context of integrals invariant under symmetry \emph{groups}.
Invariance of the integral under the action of an operator
$\mathcal{V}$ can be formulated in such a way
that $\mathcal{V}$ acts on the integral just as on the 
trivial one-dimensional representation $\mathbb{C}$, an invariant action
transforms like a complex number.

With this notion of invariance, we can construct an action from fields
which are modules of $SO_a(n)$ using the inner product integral. If the field
$\hat{\psi}$ transforms under $\hat{M}^{\mu\nu}$, then the dual object, i.e. the linear
form mapping $\hat{\psi}$ into complex numbers, has to transform under
the antipode  $S(\hat{M}^{\mu\nu})$.
The condition which the antipode of an arbitrary Hopf algebra has to fulfil is \cite{jantzen}
\begin{equation}
\label{in81}
\mathsf{m}(S\otimes 1) \Delta = \eta  \epsilon,\quad
\textrm{and}\quad \mathsf{m}(1\otimes S) \Delta = \eta \epsilon.
\end{equation}
Here $\mathsf{m}$ denotes the multiplication of two factors of a tensor
product, $\eta$ is the unit embedding $\mathbb{C}$ into $SO_a(n)$, $\Delta$ the
coproduct, and $\epsilon$ the counit (cp. \cite{f3}).
We can therefore prove the invariance of an action integral under
$SO_a (n)$. We have to verify that (we choose the convention that the dual space is
the factor on the right hand side of the inner product)
\begin{equation}
\label{in82}
(\hat{M}^{\mu\nu}\hat{\psi} , \hat{\phi})= (\hat{\psi}, S(\hat{M}^{\mu\nu})\hat{\phi}).
\end{equation}
Writing the inner product  for two $r$-forms $\psi$ and $\phi$ explicitly, we obtain the condition that (with the Hodge-dual form on the right in the inner product):
\begin{equation}
\label{in83}
\int \hspace{2mm} \big(M^{\star\mu\nu} \psi \big)\star(*\phi)=
\int \hspace{2mm}\psi\star ( S(M^{\star\mu\nu})*\phi).
\end{equation}
Note that in (\ref{in83}) the differential forms contributing to the volume element $\diff^n x$ are still
split up among the forms $\psi$ and $\phi$. 
In the following, we want to check that this condition is
fulfilled for the inner product. The check can be performed explicitly 
in the $\star$-product setting, using partial integration.

First we repeat the definition of the antipode on symmetry generators:
\begin{eqnarray}
\label{in84}
S(\hp_j)&=&-e^{-ia\hp_n}\hp_j,\qquad
S(\hp_n)=-\hp_n,\qquad
S(e^{ia\hp_n})=e^{-ia\hp_n},\nonumber\\
S(\hdi_j)&=&-e^{ia\hp_n}\hdi_j,\qquad\hspace{2mm} S(\hdi_n)=-\hdi_n+ia\hdi_j\hdi_j e^{ia\hp_n},
\\
S(\hat{M}^{rs})&=&-\hat{M}^{rs},\qquad\quad \hspace{4mm}
S(\hat{N}^l)= -\hat{N}^l e^{-ia\hp_n}-ia \hat{M}^{lk}\hp_k e^{-ia\hp_n} -ia(n-1)\hp_le^{-ia\hp_n}.\nn
\end{eqnarray}
The antipode of the coordinates $\xx^\mu$ is a
priori not
defined in our approach, since coordinates here 
are not regarded as finite translations, i.e. as elements of the $\ka$-deformed
Euclidean/Poincar\' e group, the dual
Hopf algebra of $SO_a(n)$. Therefore no coproduct is defined for
the coordinates, but formally the commutation
relations of $\xx^\mu$ with an arbitrary function can be interpreted
as a coproduct\footnote{This leads to the
same result as derived in the framework of $\ka$-deformed Euclidean/Poincar\' e group.}:
\begin{eqnarray}
\label{in85}
\xx^j \hat{f}(\xx) &=& (e^{-ia\hp_n}\hat{f}(\xx))\xx^j,\qquad\qquad\quad\longrightarrow
\hspace{2mm}\xx^j \otimes 1 - e^{-ia\hp_n}\otimes \xx^j=0,\nn\\
\xx^n \hat{f}(\xx) &=& \hat{f}(\xx)\xx^n+(ia\xx^k\hp_k\hat{f}(\xx)),\hspace{7mm} \longrightarrow\hspace{2mm}
\xx^n \otimes 1 - 1 \otimes
\xx^n-ia\xx^k\hp_k\otimes 1=0,\nn\\
S(\xx^j)&=&\xx^je^{ia\hp_n},\\
S(\xx^n)&=&\xx^n-ia\hp_k\xx^k=\xx^n-ia\xx^k\hp_k-ia(n-1).\nn
\end{eqnarray}
Regarding hermitian conjugation and antipode, there are two
particularly problematic operators  (cp. appendix \ref{hermitian}),
$\hat{N}^l$ and $\xx^n$. For further clarity we define the
$\star$-representations $S(\mathcal{V}^\star)=S(\mathcal{V})^\star$, derived 
from (\ref{in84}):
{\small
\begin{eqnarray}
\label{in86}
S(\p^\star_j)&=&-\p_j\frac{e^{-ia\p_n}-1}{-ia\p_n},\qquad
S(\p_n)=-\p_n,\qquad
S(e^{ia\p_n})=e^{-ia\p_n},\nonumber\\
S(D^\star_j)&=&-\p_j\frac{e^{ia\p_n}-1}{ia\p_n},\qquad
 S(D^\star_n)=-\frac{1}{a}\sin(a\p_n)+\frac{\p_k\p_k}{ia\p_n\p_n}(\cos(a\pat_n)-1),\nn\\
S(M^{\star rs})&=&-x^s\p_r+x^r\p_s,\\
S(N^{\star l})&=&-x^l\pat_n\frac{e^{-ia \pat_n}+1}{2} + x^n\pat_l\frac{e^{-ia \pat_n}-1}{-ia\pat_n}
+ x^l\pat_k\pat_k \frac{e^{-ia \pat_n}-1}{-2\pat_n} \nn\\&&- x^k\pat_k\pat_l
\frac{e^{-ia\pat_n} -1+ ia\pat_n}{ia\pat_n^2}+(n-1)\p_l \frac{e^{-ia \pat_n}-1}{\p_n}.\nn
\end{eqnarray}}

Comparing the $\star$-representations $S(\mathcal{V})^\star$ with the
result of  hermitian conjugation
$\mathcal{V}^\star\rightarrow\overline{\mathcal{V}^\star}$
(integrating $\mathcal{V}^\star$ under an integral
fulfilling Stokes' law, cp. appendix \ref{hermitian}), we find that they are almost identical (the definition of the antipode does not involve complex conjugation 
$i\rightarrow -i$). Of course, for this partial integration we have to employ the integral
definition involving the measure $\mu$ and the rescaling
$\p_j\rightarrow \tilde{\p}_j=\p_j+\rho_j$ (cp. \cite{f1} and appendix
\ref{hermitian}). For example:
\begin{eqnarray}
\label{in87}
\int  \mu \hspace{1mm}\big(\tilde{D}^\star_j \tilde{\psi}\big) \star \tilde{\phi} &=&\int
  \mu \hspace{1mm}\Big(\tilde{\pat}_j
\frac{e^{-ia\pat_n}-1}{-ia\pat_n}\tilde{\psi} \Big) \star  \tilde{\phi}
= \int   \mu  \tilde{\psi}\star \Big(-\tilde{\pat}_j
\frac{e^{ia\pat_n}-1}{ia\pat_n} \tilde{\phi}\Big)\nn\\
&=& \int   \mu \hspace{1mm}\tilde{\psi}\star\Big(-\tilde{D}^\star_je^{ia\pat_n} \tilde{\phi}\Big)=\int \mu \hspace{1mm}\tilde{\psi}\star\Big(S(\tilde{D}^\star_j) \tilde{\phi}\Big).
\end{eqnarray}
In this realisation of the antipode in terms of partial
integration of rescaled partial derivatives  almost
all operators can be treated in a satisfactory way. However, as in
appendix \ref{hermitian}, $\tilde{N}^{\star l}$ and $\tilde{x}^{\star n}$ again  do not fit into
the framework. The problematic piece is the factor proportional to
$(n-1)$: 
\begin{equation}
\label{in88}
S(\tilde{N}^{l})^\star =\dots+(n-1)\p_l \frac{e^{-ia \pat_n}-1}{\p_n}=(n-1)\big(-ia\p_l-\frac{a^2}{2}\p_n\p_l+\dots\big).
\end{equation}
Although we
obtain a factor proportional to $(n-1)$ from partially integrating
$\tilde{N}^{\star l}$ (from the term proportional to $x^j\tilde{\p}_j$)
\begin{equation}
\label{in89}
\tilde{N}^{\star l}\stackrel{\textrm{p.i.}}{\longrightarrow}\dots- (n-1)\pat_l
\frac{e^{-ia\pat_n} -1+ ia\pat_n}{ia\pat_n^2}=(n-1)\big(-\frac{ia}{2}\p_l-\frac{a^2}{6}\p_l\p_n+\dots\big),
\end{equation}
 this is not the right term for  $S(\tilde{N}^{l})^\star $. Changing the
 definition of $\mu$ or the rescaling $\rho_j$ to account
for the additional terms does not work, since this
would spoil the behaviour of other operators under partial
integration.

The only possibility  to obtain new terms proportional to $(n-1)$
is to fix the (rescaled) antipode $S(\tilde{N}^{l})^\star $, by introducing a derivative operator which acts on the
coordinate $x^n$. We need an asymmetrically acting operator $K$, which is
a power series in the
derivatives $\p_n$ (it must not depend on coordinates $x^\mu$ or on
$\p_j$). We define $K$ such that for all $\mathcal{V}\in SO_a(n)$
(including coordinates) the
following equation is valid (for arbitrary $r$-forms $\tilde{\psi}$ and $\tilde{\phi}$) 
\begin{equation}
\label{in90}
\int \mu \hspace{1mm}(\tilde{\mathcal{V}}^\star \tilde{\psi})\star (K
\tilde{\phi})\equiv \int \mu\hspace{1mm} \tilde{\psi}\star \Big(K\big(S(\tilde{\mathcal{V}}^\star )\tilde{\phi}\big)\Big).
\end{equation}
To simplify the subsequent calculation, we eliminate the
$\star$-product, afterwards we eliminate the measure and  the rescaling
$\tilde{\mathcal{V}}^\star \rightarrow\mathcal{V}^\star $ by
introducing the field redefinition
$\tilde{\phi}=\mu^{-\frac{1}{2}}\phi$ (according to the prescription in
\cite{f1}, $\mu^{-\frac{1}{2}}$ commutes with $K$):
\begin{equation}
\label{in91}
\int (\mathcal{V}^\star \psi) (K \phi)\equiv \int \psi \Big(K\big(S(\mathcal{V}^\star )\phi\big)\Big).
\end{equation}
The result
of the calculation below does not depend on this field
redefinition.

From (\ref{in88}) and (\ref{in89}) follows that the equation that $K$ has to satisfy reads
\begin{eqnarray}
\label{in92}
\lb K,\big(- x^n\pat_l\frac{e^{-ia \pat_n}-1}{-ia\pat_n}\big)\rb&\shouldid&
(n-1)\p_l \Big(\frac{e^{-ia\p_n}-1}{\p_n}+\frac{e^{-ia\pat_n} -1+
  ia\pat_n}{ia\pat_n^2}\Big)K,\nn\\
\Leftrightarrow\quad\frac{\p K}{\p \p_n}&\shouldid&-(n-1)\frac{-ia\pat_n}{e^{-ia
    \pat_n}-1}\big(\frac{ia\p_ne^{-ia\p_n}+e^{-ia\pat_n}
  -1}{ia\pat_n^2}\big)K,\\
\Leftrightarrow\quad K&=&c\big( \frac{-ia\pat_n}{e^{-ia\pat_n}-1}\big)^{n-1}.\nn
\end{eqnarray}
The solution is unique up to a complex multiplicative factor $c$ which
we fix $c=1$, such that $K=1+\mathcal{O}(a)$, i.e. a well-behaved commutative limit. 

This operator $K$ is the  derivative operator that
we have guessed as the result of extracting the measure $\mu$ from one of the two factors of
the $\star$-product. This means that by  constructing an action
in terms of differential forms with the Hodge-$*$ we have found an
action which is at the same time invariant under all $\mathcal{V}\in SO_a(n)$
\begin{displaymath}
(\hat{\mathcal{V}}\hat{\psi} , \hat{\phi})= (\hat{\psi},S(\hat{\mathcal{V}}) \hat{\phi}),
\end{displaymath}
since
{\small \begin{eqnarray}
\label{in94}
&&\quad\int \hspace{2mm}\Big(\tilde{\mathcal{V}}^{\star } \big(\tilde{\psi}_{\mu_1\dots\mu_r}\omega^{\mu_1}\dots\omega^{\mu_r}\big)\Big) \star
*\big(\tilde{\phi}_{\nu_1\dots\nu_r}\omega^{\nu_1}\dots\omega^{\nu_r}\big)=\nn\\
&&\qquad\qquad =\int
(\tilde{\psi}_{\mu_1\dots\mu_r}\omega^{\mu_1}\dots\omega^{\mu_r}) \star \Big( S(\tilde{\mathcal{V}}^{\star })*\big(\tilde{\phi}_{\nu_1\dots\nu_r}\omega^{\nu_1}\dots\omega^{\nu_r}\big)\Big),\nn\\
&&\Leftrightarrow\int\mu \hspace{2mm}\Big(\tilde{\mathcal{V}}^{\star } \big(\tilde{\psi}_{\mu_1\dots\mu_r}\frac{1}{r!}\omega^{\mu_1}\dots\omega^{\mu_r}\big)\Big) \star
\big(K\tilde{\phi}_{\nu_1\dots\nu_r}\frac{\epsilon^{\nu_1\dots\nu_r}_{\ \ \
  \mu_{r+1}\dots \mu_{n}}}{r!(n-r)!}\omega^{\mu_{r+1}}\dots\omega^{\mu_n}\big)=\nn\\
&&\qquad\qquad  =\int\mu\hspace{2mm}
(\tilde{\psi}_{\mu_1\dots\mu_r}\frac{1}{r!}\omega^{\mu_1}\dots\omega^{\mu_r}) \star \Big(K S(\tilde{\mathcal{V}}^{\star })\big(\tilde{\phi}_{\nu_1\dots\nu_r}\frac{\epsilon^{\nu_1\dots\nu_r}_{\ \ \
  \mu_{r+1}\dots \mu_{n}}}{r!(n-r)!}\omega^{\mu_{r+1}}\dots\omega^{\mu_n}\big)\Big),\nn\\
&&\Leftrightarrow\int \hspace{2mm}\Big(\mathcal{V}^{\star } \big(\psi_{\mu_1\dots\mu_r}\frac{1}{r!}\omega^{\mu_1}\dots\omega^{\mu_r}\big)\Big)
\big(K(\phi_{\nu_1\dots\nu_r}\frac{\epsilon^{\nu_1\dots\nu_r}_{\ \ \
  \mu_{r+1}\dots \mu_{n}}}{r!(n-r)!}\omega^{\mu_{r+1}}\dots\omega^{\mu_n})\big)=\\
&&\qquad\qquad  =\int
(\psi_{\mu_1\dots\mu_r}\frac{1}{r!}\omega^{\mu_1}\dots\omega^{\mu_r}) \Big(K S(\mathcal{V}^{\star })\big(\phi_{\nu_1\dots\nu_r}\frac{\epsilon^{\nu_1\dots\nu_r}_{\ \ \
  \mu_{r+1}\dots \mu_{n}}}{r!(n-r)!}\omega^{\mu_{r+1}}\dots\omega^{\mu_n}\big)\Big).\nn
\end{eqnarray}}
The same is valid 
for the coordinates in the definition (\ref{in85})
\begin{equation}
\label{in95}
\int  ( x^{\overrightarrow{\star }\mu}\psi)
\big(K \phi\big)=\int \psi\big(K S(x^{\overrightarrow{\star }\mu})
 \phi\big).
\end{equation}

The last step in the derivation of an invariant integral is to extract
from formulae such as (\ref{in94}) the one-forms $\omega^\mu$ and to
combine them into the volume form. We have to be careful in performing
this step, since
$N^{\star l}$ acts non-trivially on the frame one-forms (\ref{d67}). We derive the final
result in two steps: first we treat the special case of the inner
product of two functions, i.e. two zero-forms. The Hodge dual of
a function is proportional to the volume form $\diff^n x$. According to (\ref{d40a}) $\diff^n x$ transforms as 
\begin{equation}
\label{in96}
\lb \hat{N}^l, \diff^n x\rb=-ia(n-1)\diff^n x\hspace{1mm}\hp_l.\nn
\end{equation}
On the other hand
\begin{equation}
\label{in97}
S(\hat{N}^l)= -\hat{N}^l e^{-ia\hp_n}-ia \hat{M}^{lk}\pat_k e^{-ia\hp_n} -ia(n-1)\hp_le^{-ia\hp_n}.
\end{equation}
Since $\lb \hat{M}^{rs}, \diff^n x\rb =0$ and $\lb \hp_\mu,
\hom^\nu\rb=0$, we obtain
\begin{equation}
\label{in98}
S(\hat{N}^l) \diff^n x=\diff^n x\hspace{1mm}(-\hat{N}^le^{-ia\hp_n}-ia \hat{M}^{lk}\hp_k e^{-ia\hp_n}), 
\end{equation}
The term appearing at the right hand side of (\ref{in98}) is (in
$\star$-product language)
\begin{equation}
\label{in99}
-N^{\star l}e^{-ia\p_n}-iaM^{\star lk}\p^\star _k e^{-ia\p_n}=-\overline{N^{\star l}},
\end{equation}
where the bar denotes complex conjugation.
Therefore we can equivalently rewrite (\ref{in94}) for the case in which $\psi$ and
$\phi$ are two \emph{complex valued} zero-forms:
\begin{eqnarray}
\label{in100}
\int \diff^nx \big(N^{\star l}\psi\big)(K\overline{\phi} )=\int \big(N^{\star l}\psi\big)
\big(K(\overline{\phi}\hspace{1.5mm} \diff^nx )\big)&=&\int \psi \hspace{1.5mm}\Big(K
S(N^{\star l})(\phi\hspace{1.5mm}\diff^nx)\Big)\nn\\&=&-\int\diff^nx \hspace{1.5mm} \psi\hspace{1.5mm} \Big(K \overline{N^{\star l}\phi})\Big).
\end{eqnarray}
The same steps can be repeated, if $\psi$ and $\phi$ are
$r$-forms. We may commute $\omega^\mu$ with the coefficient
functions (we regard the case, in which $\omega^l$ is in the first
factor, the other case is analogous): 
{\small \begin{eqnarray}
\label{in101}
&&\int \Big(N^{\star l} \big(\omega^{\mu_1}\dots\omega^{\mu_r}\frac{1}{r!}\psi_{\mu_1\dots\mu_r}\big)\Big)
\big(K\omega^{\mu_{r+1}}\dots\omega^{\mu_n}\frac{\epsilon^{\nu_1\dots\nu_r}_{\ \ \
  \mu_{r+1}\dots \mu_{n}}}{r!(n-r)!}\phi_{\nu_1\dots\nu_r}\big)\nn\\
&&\hspace{3mm}=\int\Big(\omega^{\mu_1}\dots\omega^{\mu_r}\frac{1}{r!} \big(N^{\star l}\psi_{\mu_1\dots\mu_r}-ia(r-1)\p_l^\star \psi_{\mu_1\dots\mu_r}\big)\Big)
\big(K\omega^{\mu_{r+1}}\dots\omega^{\mu_n}\frac{\epsilon^{\nu_1\dots\nu_r}_{\ \ \
  \mu_{r+1}\dots \mu_{n}}}{r!(n-r)!}\phi_{\nu_1\dots\nu_r}\big)\nn\\
&&\hspace{3mm}=\int \diff^n x\hspace{2mm}
\big(N^{\star l}\psi_{\mu_1\dots\mu_r}\big)  \big(K\phi^{\mu_1\dots\mu_r}\big)   -ia(r-1)\int \diff^n x\hspace{2mm}\big(\p_l^\star \psi_{\mu_1\dots\mu_r}\big)\big(K\phi^{\mu_1\dots\mu_r}\big) \shouldid\nn\\
&&\shouldid\int  \big(\omega^{\mu_1}\dots\omega^{\mu_r}\frac{1}{r!}\psi_{\mu_1\dots\mu_r}\big)
\big(K S(N^{\star l})\omega^{\mu_{r+1}}\dots\omega^{\mu_n}\frac{\epsilon^{\nu_1\dots\nu_r}_{\ \ \
  \mu_{r+1}\dots \mu_{n}}}{r!(n-r)!}\phi_{\nu_1\dots\nu_r}\big)\\
&&\hspace{3mm}=\int \big(\omega^{\mu_1}\dots\omega^{\mu_r}\frac{1}{r!}\psi_{\mu_1\dots\mu_r}\big)
\Big(K \omega^{\mu_{r+1}}\dots\omega^{\mu_n}\frac{\epsilon^{\nu_1\dots\nu_r}_{\ \ \
  \mu_{r+1}\dots \mu_{n}}}{r!(n-r)!}\cdot\nn\\
&&\qquad\qquad\qquad\qquad\qquad\qquad\cdot\big(
\overline{(-N^{\star l})}\phi_{\nu_1\dots\nu_r}+ia\big((n-1)-(n-r)\big)\p_l^\star e^{-ia\p_n}\phi_{\nu_1\dots\nu_r}\big)\Big)\nn\\
&&\hspace{3mm}=\int \diff^n x\hspace{2mm}
\psi_{\mu_1\dots\mu_r}  \big(K\overline{(-N^{\star l})}\phi^{\mu_1\dots\mu_r}\big)   +ia(r-1)\int \diff^n x\hspace{2mm}\psi_{\mu_1\dots\mu_r}\big(K\p_l^\star e^{-ia\p_n}\phi^{\mu_1\dots\mu_r}\big). \nn
\end{eqnarray}}
Partially integrating the term proportionally to $(r-1)$, the result
for complex valued forms is:
\begin{equation}
\label{in102}
\int \diff^n x\hspace{2mm}
\big(N^{\star l}\psi_{\mu_1\dots\mu_r}\big)  \big(K\overline{\phi^{\mu_1\dots\mu_r}}\big) =-\int \diff^n x\hspace{2mm}
\psi_{\mu_1\dots\mu_r}  \big(K\overline{N^{\star l}\phi^{\mu_1\dots\mu_r}}\big).
\end{equation}
This identity is valid by partial integration and taking into account
the action on the volume element and  the commutation relation with
$K$. From an abstract definition of inner
product we have derived a hermitian representation of $N^{\star l}$. More importantly, (\ref{in102}) shows 
that an action defined in terms of forms is invariant under $N^{\star l}$.

All other operators $M^{\star rs}$ and the derivatives $D_\mu^\star $ and
$\p_\mu^\star $ (no tilde) can be treated analogously. The discussion of these operators is straightforward since they commute with $K$ and with the
volume element $\diff^n x$ and they   be partially integrated without harm
(since $\mu$ has been eliminated).

The integral just defined is obviously not cyclic, since from the outset we
have discussed an asymmetric setting: the $\star$-product is not
commutative and therefore it matters whether the Hodge-dual form is
in the first or in the second place of the inner product. For the Hopf
algebra setting, this however is essential: the order in the inner
product \emph{must not} be reversible, since the module space and its
second dual space, i.e. the dual of the dual space, are not
identical. We recall \cite{f3} that the square of
the antipode is not the identity:
\begin{displaymath}
S^2(\hat{N}^l)= \hat{N}^l+ia(n-1)\hat{\partial}_l\ne \hat{N}^l.
\end{displaymath}
The generator $N^{\star l}$ acts in
different ways on a space and its second dual. Therefore it is clear
that in formulae such as (\ref{in94}) we cannot simply partially integrate once more
to obtain the action on the second dual space. The construction
of the bidual space has to be redone from  scratch. We will not
repeat the calculations (via partial integration as above), but they
give the following result  ($\psi$ and $\phi$ arbitrary $r$-forms)
\begin{equation}
\label{in103}
\int \big(S(N^{\star l})\psi\big)  \big(K\phi\big) =\int\psi \big(KS^2(N^{\star l})\phi\big).
\end{equation}
This indeed gives the correct result for the algebraic expression of
the square of the antipode. Because of this property, derivative
operators such as $K$ generally occur for traces for general Hopf algebras \cite{jantzen}. The integral
defined with such an operator is called the
\emph{quantum trace}.

We have not been able yet to fully understand the usefulness of the quantum trace. The integral over a field $\int
\psi(x)$ can sensibly analysed in this way using $*1\sim
\mu\textrm{d}^nx$. The product of several fields has to be discussed
 with great care, as usual in the differential form setting.
The operator $K$ can be partially
integrated onto the other form: 
\begin{equation}
\label{in52a}\int \diff^n x\hspace{2mm}\psi  \Big(\big(\frac{-ia\p_n}{e^{-ia\p_n}-1}\big)^{n-1}\phi\Big)
=\int\diff^n x\hspace{2mm} \Big(\big(\frac{ia\p_n}{e^{ia\p_n}-1}\big)^{n-1} \psi\Big) \phi.
\end{equation}

The most pressing problem of the quantum trace is that a priori it does  not allow to
formulate gauge invariant actions from gauge covariant Lagrangians,
since it is not cyclic (cp. appendix \ref{cyclic}). Still it is possible to
formulate a gauge-covariantised version of the quantum trace, since
the derivative $\p_n$ with an ordinary Leibniz rule can be gauged
(cp. the procedure in \cite{f2}). Similarly,
functions of $\p_n$ (such as $K$) can be gauged as well by gauging
every single derivative and $\star$-multiplying the
gauge-covariantised derivatives. Therefore a gauge-covariantised
version of $K$ is possible, however, it is difficult to see how a
covariantised $K$ still might provide an $SO_a(n)$-invariant integral. The
only possiblity would be to choose a particular gauge for the gauge
potential, i.e. the gauge potential corresponding to $\p_n$ identical
to zero. 

The upshot of this discussion is that we have presented a new
definition for an integral on the $\ka$-deformed spacetime. It is
definitely invariant under the deformed symmetry and has a well-defined geometric
interpretation. However, it does not yet have all properties that
are desirable for a physical integral. Currently, we still have to
choose between formulations of the integral which are either not
invariant under symmetry transformations (at least at face value) or
not gauge invariant (at least at face value). This is the same
conclusion as the authors of \cite{gac2} have recently drawn
w.r.t. the cyclic integral that  we discuss for reference in
the appendix.

\begin{appendix}

\section{Cyclic integral}
\label{cyclic}
 An integral may be formulated as a linear map of the coordinate
 algebra $\mathcal{A}_{\hat{x}}$
into the number field on which it is defined:
\begin{equation}
\label{in3}
\int: \mathcal{A}_{\hat{x}}\longrightarrow \mathbb{C},
\end{equation}

\begin{equation}
\label{in4}
\int (c_1\hat{\psi}+c_2\hat{\phi})= c_1\int\hat{\psi}+c_2\int\hat{\phi},\qquad
\forall \hat{\psi}, \hat{\phi} \in \mathcal{A}_{\hat{x}},\hspace{2mm} c_i \in \mathbb{C}.
\end{equation}

In addition we demand the trace property in this first section of the appendix:
\begin{equation}
\label{in5}
\int\hat{\psi}\hat{\phi}=\int\hat{\phi} \hat{\psi}. 
\end{equation}
The trace property implies that the integral is  cyclic
$\int\hat{\psi}\hat{\phi}\hat{\chi}=\int\hat{\phi}\hat{\chi}
\hat{\psi}$. 
The integral on the abstract algebra has to be realised in terms of an
integral over commutative space, such as the Lebesgue
integral. Therefore we need a realisation of the integral in the
$\star$-product formalism to perform integration explicitly. 

An essential property of the integral is that it allows the use of
Stokes' theorem; therefore there are also additional restrictions on
the space of allowed functions (their total derivatives must
vanish).

Provided that all derivatives, which arise due to an expansion of the  $\star$-product,
could be eliminated by partial integration at every order,
 it reduces to point-wise multiplication (\ref{in5}).  Such a
 procedure is possible for the  Moyal-Weyl
$\star$-product, but not for an  $x$-dependent
$\star$-product.  For example the partial integration in the
case of $\ka$-deformed space delivers in first order:
\begin{eqnarray}
\label{in5b}
&&\int\textrm{d}^n x  \hspace{1mm}\frac{ia}{2}\big((\pat_n\psi(x))(x^j\pat_j
\phi(x))-(x^j\pat_j \psi(x))(\pat_n \phi(x))\big)\nn\\
&&\qquad\qquad\stackrel{\textrm{part. int.}}\longrightarrow  \frac{ia}{2}\int\textrm{d}^n x \hspace{1mm}\Big(\psi(x) \phi(x) -(n-1) (\pat_n\psi(x)) \phi(x)\Big).
\end{eqnarray}

It has been shown in the framework of deformation quantisation of Poisson manifolds \cite{shoikhet2}
that it is always possible to define a
measure function  $\mu(x)$ (which is part of the volume element)
 such that the integral of two functions
multiplied with the $\star$-product is cyclic. 
This has been shown in \cite{dietz} in a constructive way for quantum spaces. The
measure for the  $\ka$-deformed spacetime has been discussed first in
\cite{f1} and then in \cite{syko2} from the deformation quantisation
perspective. For an $x$-dependent $\star$-product
$\theta^{\rho\si}(x)$ the measure function $\mu(x)$ has to fulfil the condition:
\begin{equation}
\label{in5c}
\pat_\rho\big(\mu(x)\theta^{\rho\si}(x)\big)=0.
\end{equation}
This statement is an all-orders statement (cp. the discussion in \cite{gac2}). For $\ka$-deformed spacetime (\ref{in5c}) entails the following conditions on $\mu(x)$:
\begin{equation}
\label{in7}
\pat_\rho \big(\mu(x)a(\de^\rho_nx^\si-\de^\si_nx^\rho)\big)=0
\quad\Rightarrow\quad \partial_n \mu(x) =0, \qquad x^j \partial_j \mu (x)= -(n-1)\mu(x).
\end{equation}
Examples of measures $\mu(x)$ fulfilling
(\ref{in7}) are 
\begin{equation}
\label{in8}
\mu_1(x)=\Big(\prod_{i=1}^{n-1}x^i\Big)^{-1},\quad
\mu_2(x)=\Big(\sum_{i=1}^{n-1}(x^ix^i)\Big)^{-\frac{(n-1)}{2}},\quad
\mu_k(x)=\Big(\sum_{i=1}^{n-1}(x^i)^k\Big)^{-\frac{(n-1)}{k}},
\forall k \in \mathbb{N}.
\end{equation}
If $\mu(x)$ is given, the integral over the $\star$-product of two functions has the trace property:
\begin{equation}
\label{in6}
\int\textrm{d}^n x \hspace{1mm}\mu(x)\hspace{1mm}(\psi(x)\star \phi(x)) =\int\textrm{d}^n x\hspace{1mm} \mu(x)\hspace{1mm}(\phi(x)\star \psi(x)) =\int \textrm{d}^n x \hspace{1mm} \mu(x)\hspace{1mm}\psi(x) \phi(x). 
\end{equation}

The measure $\mu(x)$ allows to eliminate any one
of the $\star$-products from the $\star$-product of several
functions, because of associativity.
This allows to cyclically permute under the integral an arbitrary number of
$\star$-multiplied functions
\begin{equation}
\label{in10}
\int\textrm{d}^n x\hspace{1mm} \mu(x)\hspace{1mm}(\psi_1(x)\star \dots
\star \psi_k(x))=\int\textrm{d}^n x\hspace{1mm} \mu(x)\hspace{1mm}(\psi_k(x)\star
\psi_1(x)\star \dots \star \psi_{k-1}(x)).
\end{equation}
Therefore gauge covariant Lagrangians lead to gauge invariant actions.

\section{Hermitian derivative operators}
\label{hermitian}

Conjugation $ ^{\dagger}$ can be defined on the $\ka$-deformed coordinate algebra  and also on its symmetry
Hopf algebra $SO_a(n)$  as a formal
involution. We demand:
\begin{itemize}
\item Consistency with the algebraic structure $(\lb \mathcal{V},\mathcal{W}\rb-\mathcal{U})^\dagger=0$, if $\lb \mathcal{V},\mathcal{W}\rb-\mathcal{U}=0$.
\item Complex conjugation for numbers.
\item Conjugation is an involution $(\mathcal{V}\mathcal{W})^\dagger=\mathcal{W}^\dagger\mathcal{V}^\dagger$.
\end{itemize}
An operator is hermitian if $ \mathcal{V}^\dagger=\mathcal{V}$.
For a well-defined commutative limit, coordinates
should be hermitian and derivatives anti-hermitian. We can calculate
the conjugation properties of the symmetry generators from their 
representation in terms of $\xx^\mu$ and $\hp_\mu$:
\begin{eqnarray}
\label{in11b}
(\xx^\mu)^\dagger&=&\xx^\mu,\qquad
(\hat{\partial}_n)^\dagger=-\hat{\partial}_n,\qquad
(\hat{\partial}_j)^\dagger=-\hat{\partial}_j,\nn\\
(\hat{D}_j)^\dagger&=&\big(\hp_j e^{-ia\hp_n}\big)^\dagger=\big( e^{ia\hp^\dagger_n}\hp_j^\dagger\big)=-\hat{D}_j,\nn\\
(\hat{D}_n)^\dagger&=&\Big(\frac{1}{a}\sin(a\hp_n)+\frac{ia}{2}\hp_k\hp_k
e^{-ia\hp_n}\Big)^\dagger=-\hat{D}_n,\\
(\hat{M}^{rs})^\dagger&=&-\hat{M}^{rs},\qquad\qquad
(\hat{N}^l)^\dagger=  -\hat{N}^l.\nn
\end{eqnarray}
Thus, formal conjugation can be defined
consistently in the abstract algebra. In addition we need to check
the
conjugation behaviour of the $\star$-representations. Here derivative
operators should be conjugated in a concrete sense, using hermitian
conjugation implemented by partial
integration under the integral\footnote{The notion of selfadjointness
  requires careful definitions of the domain  of the
  operators.}.  
Thus, we call a derivative operator in the representation
$\mathcal{V}^\star $ hermitian 
if 
\begin{equation}
\label{in13}
\int\textrm{d}^n x\hspace{1mm} \mu\hspace{1mm}\overline{\psi}\star
\mathcal{V}^\star \phi =  \int\textrm{d}^n x\hspace{1mm}
\mu\hspace{1mm}\overline{\mathcal{V}^\star \psi}\star \phi, 
\end{equation}
under partial integration. For the two simplest derivative operators
$\pat_n^\star $ and $\pat^\star _j$  we obtain that although $\hp_\mu^\dagger=-\hp_\mu$
 (and $\overline{\pat_n^\star }=-\pat_n^\star $) 
\begin{equation}
\label{in14}
\int \textrm{d}^n x\hspace{1mm} \mu\hspace{1mm} \overline{\psi}\hspace{1mm}(\partial_i^\star 
\phi)  \stackrel{p.i.}{\rightarrow}  -\int \textrm{d}^n x\hspace{1mm} \mu\hspace{1mm} \overline{\partial_i^\star  \psi}\hspace{1mm} \phi-\int \textrm{d}^n x\hspace{1mm} \partial_i\mu\hspace{1mm} \overline{\frac{e^{ia\p_n}-1}{ia\p_n} \psi}\hspace{1mm}  \phi.
\end{equation}
The derivative $\pat_j^\star $ is not anti-hermitian,
since it acts on the measure $\mu$.
The solution to this problem \cite{f1} is a rescaling of the derivative $
\pat_j\rightarrow \tilde{\pat}_j=\pat_j+\rho_j=\pat_j+\frac{(\partial_j\mu)}{2\mu}$. 
The rescaling factor $\rho_j$ inherits from $\mu$ the 
properties:
\begin{equation}
\label{in16}
x^l\partial_l \rho_j=-\rho_j \quad \textrm{and}\quad \partial_n
\rho_j =0.
\end{equation}
For the choices of $\mu$ presented in (\ref{in8}), we
would obtain:
\begin{equation}
\label{in16a}
\rho_j(\mu_1)=-\frac{1}{2x^j},\quad\rho_j(\mu_2)=-\frac{n-1}{2}\frac{x^j}{\sum_{i=1}^{n-1}x^ix^i},\quad\rho_j(\mu_3)=-\frac{n-1}{2}\frac{(x^j)^{k-1}}{\sum_{i=1}^{n-1}(x^i)^k}.  
\end{equation}
However, it is not necessary to specify a particular
form  for $\mu$ or 
for $\rho_j$. 

With the rescaled derivative $\tilde{\pat}_j$, anti-hermitian derivative
operators can be constructed such as $\tilde{\partial}^\star _j$:
\begin{equation}
\label{in17}
{\tilde \partial}^\star _j = (\partial_j
+\rho_j)\frac{e^{ia\partial_n}-1}{ia\partial_n}. 
\end{equation}
This derivative operator $\tilde{\partial}^\star _j$ is anti-hermitian in
the sense of (\ref{in13}). Similarly, $D^\star _\mu$ are rescaled to be anti-hermitian in the sense of (\ref{in13}):
\begin{eqnarray}
\label{in19}
D^\star _j&\longrightarrow&{\tilde D}^\star _j=(\partial_j
+\rho_j)\frac{e^{-ia\partial_n}-1}{-ia\partial_n},\nonumber\\
D^\star _n&\longrightarrow&{\tilde D}^\star _n=\frac{1}{ia\partial_n^2}(\partial_k
+\rho_k)(\partial_k
+\rho_k)(\cos(a\partial_n)-1)+\frac{1}{a}\sin(a\partial_n).
\end{eqnarray}

The rescaling with $\rho_j$ is algebraically consistent \cite{f1}: $\lb(\p_j+\rho_j),
x^\mu\rb=\de_j^
\mu$ is unchanged and also
$\lb(\p_i+\rho_i),(\p_j+\rho_j)\rb=0$. Thus, the rescaling can be lifted  into
the abstract algebra. Then however, the representation of all operators
$M^{\star \mu\nu}\rightarrow
\tilde{M}^{\star \mu\nu}$ and $x^{\star }\rightarrow
\tilde{x}^{\star \mu}$ has to be changed as well: 
\begin{eqnarray}
\label{in19a}
{\tilde N}^{\star l}&=&x^l\pat_n \frac{e^{ia \pat_n}-1}{2}+ x^l\tilde{\pat}_j\tilde{\pat}_j \frac{e^{ia \pat_n}-1}{2\pat_n} - x^n\tilde{\pat}_l
\frac{e^{ia\pat_n} -1}{ia\pat_n}- x^j\tilde{\pat}_j\tilde{\pat}_l
\frac{e^{ia\pat_n} -1- ia\pat_n}{ia\pat_n^2},\nonumber\\
{\tilde M}^{\star rs}&=&x^s\tilde{\pat}_r-x^r\tilde{\pat}_s, \quad
{\tilde x}^{\star n}={\tilde x}^{\overrightarrow{\star }n}=x^n -x^k\frac{\tilde{\pat}_k}{\pat_n}\big(\frac{ia\pat_n}{e^{ia\pat_n}-1}-1\big), \quad
{\tilde x}^{\star j}={\tilde x}^{\overrightarrow{\star }j}=x^{\star j}.\nn
\end{eqnarray}
The notation ${\tilde x}^{\overrightarrow{\star }\mu}$ denotes a coordinate
multiplied from the left to another function.  

Unfortunately, even including the action of
$\tilde{N}^{\star l}$ on $\diff^n x$,
$\tilde{N}^{\star l}$ is not anti-hermitian in the sense of
(\ref{in13}). The
problematic piece is the one
 proportional to
$x^j\tilde{\pat}_j\tilde{\pat}_l$; this arises due to the
representation of ${\tilde x}^{\star n}$ which is not defined as a
hermitian quantity in this approach. We have made several attempts to
cure this problem, including the definition of the opposite algebra,
acting from the right, or an interpretation of hermitian conjugation
for coordinates as changing from  left to right multiplication. But in
none of these attempts, the hermitian conjugation of $\tilde{N}^{\star l}$
is fully satisfactory. With the cyclic integral of appendix
\ref{cyclic}, we cannot find a fully consistent definition of
hermitian operators. However, we have shown in the main part of this
note, that with the quantum trace also this problem can be handled.

\end{appendix}
\section*{Acknowledgements}
I am grateful to Marija Dimitrijevi\'c, Larisa Jonke, Harold Steinacker and Julius Wess for many
fruitful discussions, contributing to the results of this note. I am
grateful to Marija for careful proofreading.


\end{document}